\documentstyle[aps,prl,epsfig,multicol]{revtex}

\def\twobeone{\end{multicols}
\vskip.6pc \noindent \vrule width3.375in height.2pt depth.2pt
\vrule depth0em height1em\hfill \vskip.6pc}

\def\onebetwo{\vskip.6pc \indent \hfill\vrule depth1em height0pt
\vrule width3.375in height.2pt depth.2pt \vskip.6pc
\begin{multicols}{2}\noindent}

\begin{document}

\title{Magnetic properties of an SU(4) spin-orbital chain}

\author{Shi-Jian Gu and You-Quan Li}
\address{Zhejiang Institute of Modern Physics,
Zhejiang University, Hangzhou 310027, P.R. China}
\date{ Received 23 February 2002}
\maketitle

\begin{abstract}
In this paper, we study the magnetic properties of the
one-dimensional SU(4) spin-orbital model by solving its Bethe
ansatz solution numerically. It is found that the magnetic
properties of the system for the case of $g_t=1.0$ differs from
that for the case of $g_t=0.0$. The magnetization curve and
susceptibility are obtained for a system of 200 sites. For
$0<g_t<g_s$, the phase diagram depending on the magnetic field and
the ratio of Land\'e factors, $g_t/g_s$, is obtained. Four phases
with distinct magnetic properties are found.

\end{abstract}

\begin{multicols}{2}

Recently, much attention has been focused on strongly correlated
electrons with orbital degrees of freedom
\cite{YTokura00,AMOles00,AJoshi99,CItoi00,YLLee00,YYamashita00,BFrischmuth99}
due to progress in experimental studies of transition metal and
rare earth compounds such as LaMnO$_3$, CeB$_6$ and TmTe. Examples
of spin-orbital systems in one dimension include
quasi-one-dimensional
tetrahis-dimethylamino-ethylene(TDAE)-C$_{60}$\cite{DPArovas95},
artificial quantum dot arrays\cite{AOnufriev99} and degenerate
chains in Na$_2$Ti$_2$Sb$_2$O and Na$_2$V$_2$O$_5$
compounds\cite{EAxtell97,SKPati98,MIsobe96}. In these systems, the
low-lying electron states have orbital degeneracy in addition to
the spin degeneracy. This may result in various interesting
properties associated with the orbital degrees of freedom in Mott
insulating phases. For example, the magnetic ordering is
influenced by the orbital structure which may change with
pressure, or the magnetization is a nonlinear function of
magnetic field even in the case of an isotropic exchange
interaction of $S_i\cdot S_j$ type \cite{KIKugel82}

The spin model with orbital degeneracy in a magnetic field was
studied by means of effective field theories\cite{YLLee00}
recently. It has been shown that there exists critical behavior
under magnetic field. For a nonperturbative understanding of the
magnetic properties of the model, the Bethe-ansatz solution of the
SU(4) model is applicable when the exterior magnetic field is
applied since the total $z$ component of spin and orbital is a
good quantum number. Although Ref. \cite{YYamashita00} studied the
magnetic properties in terms of the Bethe-ansatz method, their
result involves only a special case because the Land\'e factor was
not taken into account. In present paper, we study the model in
the presence of an exterior field by solving the Bethe-ansatz
equation numerically meanwhile taking account of the Land\'e $g$
factor. We obtain a rich phase diagram in comparison to what was
obtained in \cite{YYamashita00}. Our results shed new light on the
understanding of more realistic systems. The quantum phase
transition(QPT)\cite{SLSondhi97} concluded from such a system is
speculated to be found in experiments.

The one-dimensional quantum  spin 1/2 system with twofold orbital
degeneracy is modelled by\cite{KIKugel73,YQLi98}
\begin{equation}
{\cal H}=\sum_{j=1}^N\left[\left(2T_j\cdot
T_{j+1}+\frac{1}{2}\right) \left(2S_j\cdot
S_{j+1}+\frac{1}{2}\right)-1\right]. \label{eq:Hamiltonian}
\end{equation}
where the coupling constant is set to unity. The Hamiltonian
(\ref{eq:Hamiltonian})  has not only SU(2)$\times$SU(2) symmetry,
but also an enlarged SU(4) symmetry\cite{YQLi98}. It was solved by
the Bethe-ansatz approach, the obtained secular equation reads
\cite{BSutherland,YQLi99}:
\begin{eqnarray}
2\pi I_a=&& N\theta_{-1/2}(\lambda_a)
 +\sum_{b=1}^M \theta_1(\lambda_a-\lambda_b)
   \nonumber\\
&&+\sum_{c=1}^{M'}\theta_{-1/2}(\lambda_a-\mu_c),
 \nonumber \\
2\pi J_a=&&\sum_{b=1}^M\theta_{-1/2}(\mu_a-\lambda_b)
 +\sum_{c=1}^{M'}\theta_1(\mu_a-\mu_c)
  \nonumber \\
&& +\sum_{d=1}^{M''}\theta_{-1/2}(\mu_a-\nu_d),
  \nonumber \\
2\pi
K_a=&&\sum_{b=1}^{M'}\theta_{-1/2}(\nu_a-\mu_b)+\sum_{c=1}^{M''}\theta_1(\nu_a-\nu_c).
\label{eq:BAE}
\end{eqnarray}
where $\theta_\rho(x)=-2\tan^{-1}(x/\rho)$. The quantum numbers
$\{I_a, J_a, K_a\}$ specify a state in which there are $N-M$
number of sites in the state $|\underline \uparrow\rangle$, $M-M'$
in $|\overline\uparrow\rangle$, $M'-M''$ in $|\underline
\downarrow\rangle$, and $M''$  in $|\overline \downarrow\rangle$.
The $\lambda, \mu$ and $\nu$ are rapidities related to the three
generators of the Cartan subalgebra of the SU(4) Lie algebra. The
coupled transcendental equations (\ref{eq:BAE}) determine the
energy spectrum of present model. To solve the Bethe ansatz
equation numerically, one needs to take into account of the
following propositions:
\begin{itemize}
\item  Given a $N$, Max$(M)=3N/4$, Max$(M')=N/2$, Max$(M'')=N/4$, and
$N-M \geq M-M' \geq M'-M'' \geq M''$;
\item $\{I_a, J_a, K_a\}$ are consecutive integers (or half integers)
arranged symmetrically around the zero, that are integers or half
integers depending on whether $N-M-M', M-M'-M'', M'-M''$ are odd
or even respectively.
\end{itemize}
The latter comes from the logarithm and the former is due to the
property of the Young tableau. Once Eqs. (\ref{eq:BAE}) are
solved, the energy in the absence of exterior field is evaluated
by
\begin{equation}
E_0(M, M', M'')=-\sum_{a=1}^M\frac{1}{1/4+\lambda_a^2}.
\end{equation}

As the $z$ component of total spin and that of total orbital are
given by $S_z=N/2-M'$ and $T_z=N/2-M+M'-M''$, the magnetization
$M_z=g_sS_z+g_tT_z$ reads
\begin{equation}
M_z=\frac{N}{2}(g_s+g_t)-M g_t-M'(g_s -g_t)-M'' g_t,
\label{EQ:MAG}
\end{equation}
where $g_s$ and $g_t$ are Land\'e $g$ factors for spin and
orbital, respectively. In the presence of the magnetic field $H$
the energy becomes
\begin{equation}
E(H, M, M', M'')=E_0 - H M_z. \label{eq:TEnergy}
\end{equation}
In the absence of a magnetic field, the lowest-energy state is a
SU(4) singlet where $M''=N/4$, $M'=N/2$ and $M=3N/2$. In the
presence of a magnetic field, however, level crossing occurs and
the state with $M''=N/4$, $M'=N/2$ and $M=3N/2$ is no longer the
ground state. Therefore the values $M, M', M''$ for the lowest
state depend on the magnitude of the applied magnetic field $H$.
By analyzing the level crossing between the lowest-energy state
and the others in terms of (\ref{eq:TEnergy}), we are able to
calculate the magnetization by Eq. (\ref{EQ:MAG}).

\begin{figure}
\setlength\epsfxsize{75mm} \epsfbox{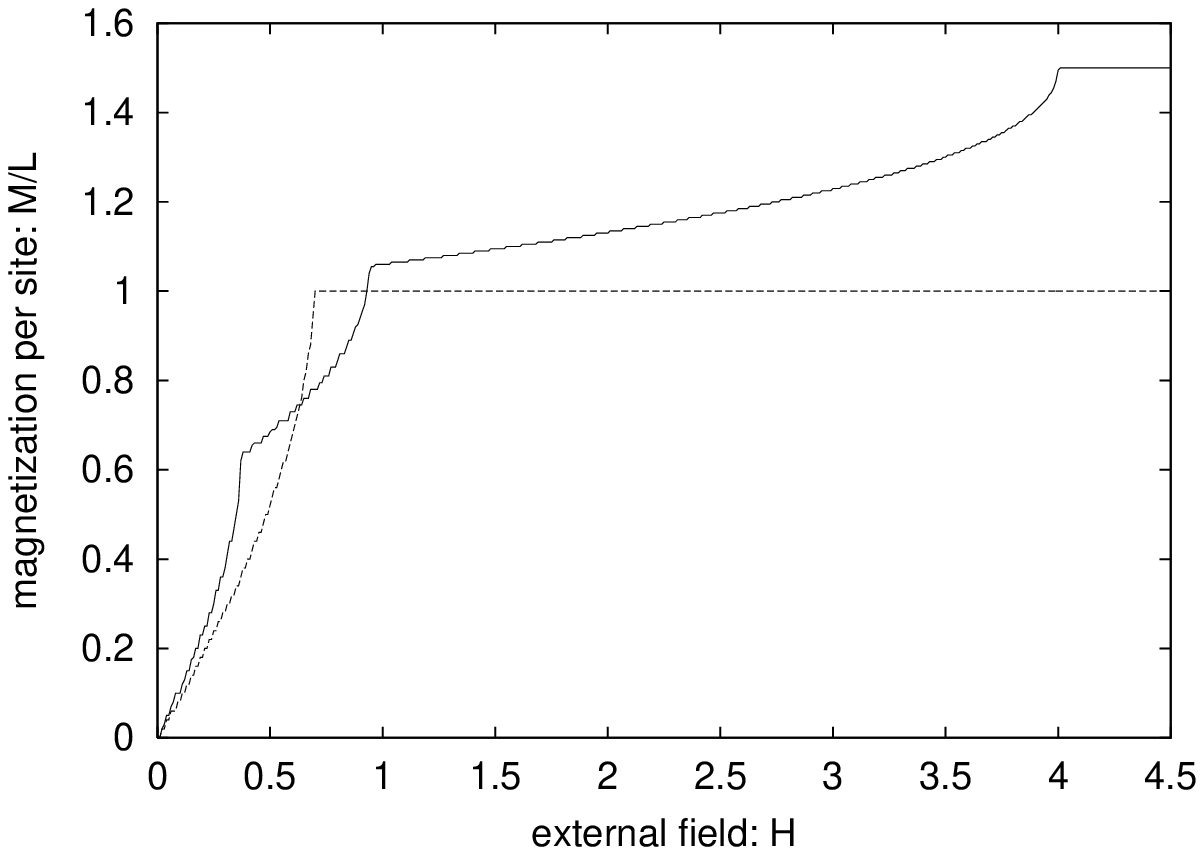}
\caption{The
magnetization of SU(4) spin-orbital chain versus external field
for 200-site system.
The solid line corresponds to $g_t=1.0$
and the dashed line corresponds to $g_t=0.0$.}
\label{FIGURE_TMAG}
\end{figure}

We set $g_s=2.0$ and observe various situations for different
values of $g_t$. The dependence of magnetization versus the
magnetic field $H$ is plotted in Fig. \ref{FIGURE_TMAG} for the
cases of $g_t=0.0$ and $g_t=1.0$ respectively. Clearly, the
magnetic properties of these two cases are quite different. We
find that there exist three critical points in the magnetization
process, of which we will give interpretations on the base of
symmetry analysis, the evolution of the Young tableau as well as
the physics picture by means of the mean-field analysis later on.

For an easier understanding, we first give an explanation by means
of the Young tableau for 8-site system. In Table I, we present the
whole 15 different multiplets, their energy in the absence of a
magnetic field, and the corresponding magnitudes of magnetization.
Each multiplet corresponds to a distinct Young tableau.
\twobeone
\begin{table}
\caption{The ground state energy $E_0$ in vanished exterior field,
the z-component of the total spin $S_z$,  that of the total
orbital $T_z$, and magnetization $M_z$ of all possible
configuration of an eight-site system}
\begin{tabular}{c|c|c|c|c|c|c|c|c|c|c|c|c|c|c|c}
Multiplet & [2$^4$] & [3221] & [3311] & [332] & [4211]
 & [422] & [431] & [44] & [51$^3$] & [521] & [53]
 & [511] & [62] & [71] & [8]
 \\ \hline
 $E_0$ & -14.92 & -13.73 & -13.84 & -13.41 & -12.76 & -12.85 & -12.04 &
 -11.30 & -10.83 & -9.95 & -10.25 & -7.41 & -7.99 & -4.00 & 0\\ \hline
$T_z$ & 0 & 1 & 0 & 1 & 1 & 2 & 1 & 0 & 2 & 2 & 1 & 3 & 2 & 3 & 4 \\ \hline
$S_z$ & 0 & 1 & 2 & 2 & 2 & 2 & 3 & 4 & 2 & 3 & 4 & 3 & 4 & 4 & 4 \\ \hline
$M_z(g_t=0)$ & 0 & 2 & 4 & 4 & 4 & 4 & 6 & 8 & 4 & 6 & 8 & 6 & 8 & 8 & 8 \\ \hline
$M_z(g_t=1)$ & 0 & 3 & 4 & 5 & 5 & 6 & 7 & 8 & 6 & 8 & 9 & 9 & 10 & 11 & 12 \\
\end{tabular}
\end{table}
\onebetwo
The application of an exterior field will make each energy level
corresponding to a multiplet that carries out a SU(4) irreducible
representation split into Zeemann sublevels. If $g_t=0.0$, the
energy level of an SU(4) singlet labelled by the Young tableau
$[2^4]$ will cross with the sublevel of the highest weight state
labelled by the Young tableau $[3 3 1 1]$ when the applied
exterior field reaches the value of $0.27$. If the exterior field
increases further this state becomes the ground state. The
magnitude of magnetization plays the role of the slope of energy
curves determined by Eq.(\ref{eq:TEnergy}). When $H=0.635$ its
energy level will cross with that of the state labelled by Young
tableau $[4 4]$. For $H>0.635$ the spins are totally polarized and
the magnetization is saturated.
\[
\setlength{\unitlength}{3mm}
\begin{picture}(18,5)(3,0)\linethickness{0.6pt}
\put(0,1){\line(1,0){2}}\put(0,2){\line(1,0){2}}\put(0,3){\line(1,0){2}}
 \put(0,0){\line(1,0){2}}\put(0,0){\line(0,1){4}}
  \put(0,4){\line(1,0){2}}\put(2,0){\line(0,1){4}}
   \put(1,0){\line(0,1){4}}
\put(3,1.5){\vector(1,0){3}}\put(3,2){$_{H> 0.27}$}
\put(7,0){\line(1,0){1}}\put(7,0){\line(0,1){4}}\put(7,1){\line(1,0){1}}
 \put(8,0){\line(0,1){4}}\put(9,2){\line(0,1){2}}\put(10,2){\line(0,1){2}}
  \put(7,4){\line(1,0){3}}\put(7,3){\line(1,0){3}}\put(7,2){\line(1,0){3}}
\put(11.5,1.5){\vector(1,0){3}}\put(11.5,2){$_{H> 0.635}$}
\put(16,2){\line(0,1){2}}\put(17,2){\line(0,1){2}}\put(18,2){\line(0,1){2}}
 \put(19,2){\line(0,1){2}}\put(20,2){\line(0,1){2}}
  \put(16,2){\line(1,0){4}}\put(16,3){\line(1,0){4}}\put(16,4){\line(1,0){4}}
\end{picture}
\]
When $g_t=g_s=1$, the level crossing occurs to the highest weight
states labelled by Young tableaux $[3311]$, $[332]$, $[422]$,
$[44]$, $[53]$, $[62]$, $[71]$, and $[8]$ for the 8-site system.

\begin{figure}
\setlength\epsfxsize{75mm} \epsfbox{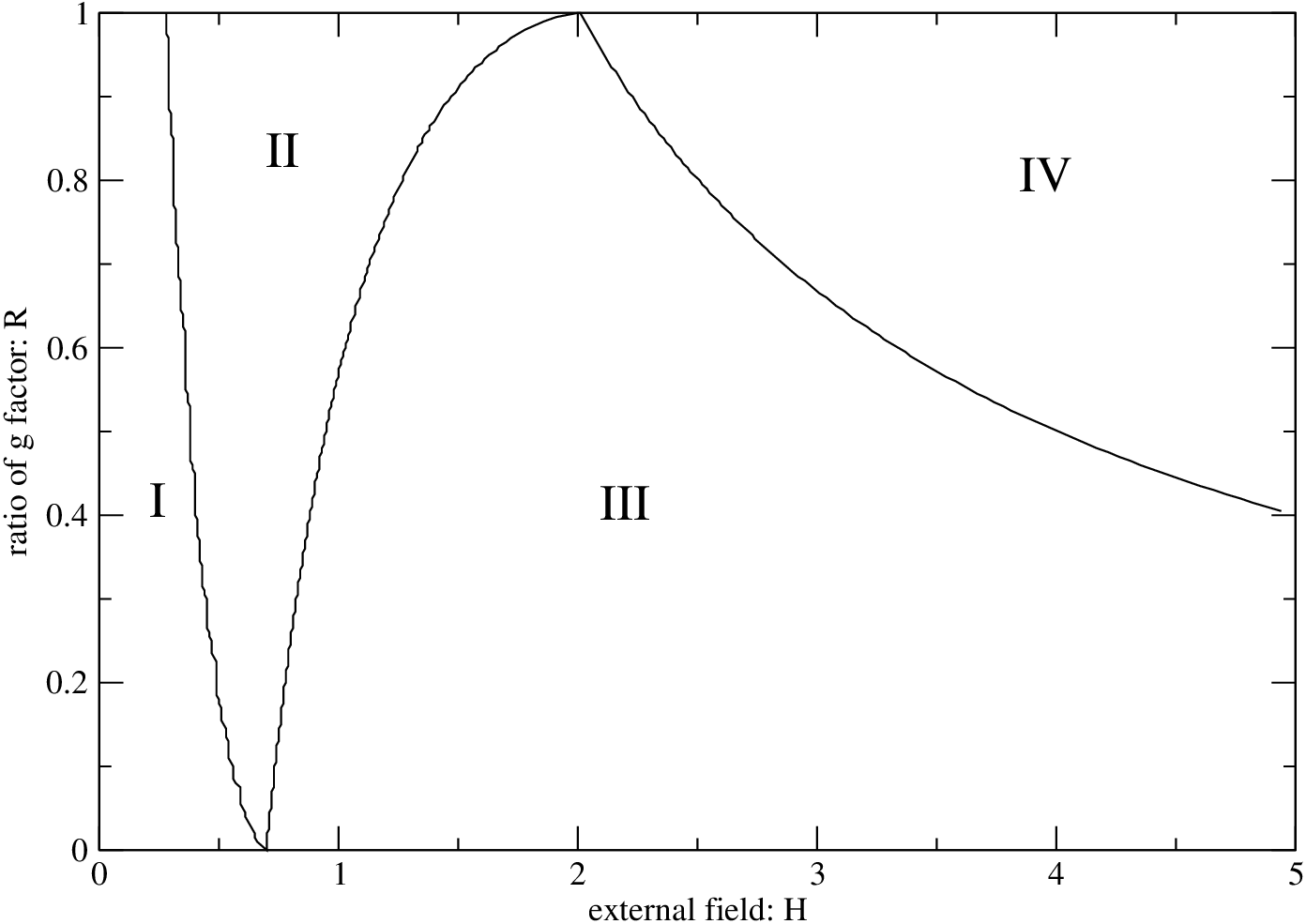} \caption{Phase
diagram of magnetic field, $H$ and the ratio of Land\'e $g$
factor, $R=g_t/g_s$. There exist separated phases: I. (SSP-WOP),
II. (SSP-OAP), III. (SF-OP), and IV. (SF-OF) in the region of
$0<R<1$. The dashed line in the middle of the figure corresponds
to $g_t=1.0$. } \label{FIGURE_PHASE}
\end{figure}

It is clear that the exterior field only breaks the spin SU(2)
symmetry and the orbital SU(2) symmetry remains when $g_t=0$. The
magnetization process is similar to that of traditional Heisenberg
model but the intermediate state are those labelled  by the Young
tableau
 $[m_1, m_2, m_3, m_4]$ with $m_1=m_2$ and $m_3=m_4$,
 i.e., $M'=2M''$,
$N+M'=2M$. This implies that the evolution of Young tableau caused
by the exterior field is realized by moving a couple of boxes
lying in the third and the fourth row to the first and the second
row. Unlike in the transitional Heisenberg model where an
increasing exterior field flips the down spins to up spins one by
one, it always flips a pair of parallel spins that constitutes an
orbital singlet. As a result, present spin-orbital model presents
larger magnetic susceptibility meanwhile it undergoes two phases
({\it i.e.,} phase I and III in Fig. \ref{FIGURE_PHASE}) versus
the exterior field.

Because the orbital SU(2) symmetry is also broken for nonvanishing
$g_t$, the magnetization process becomes complicated when
$0.0<g_t<2.0$. The whole process undergoes four regimes as long as
$g_t < g_s$. The first regime consists of strong spin polarization
and weak orbital polarization (SSP-WOP) processes, where the level
crossing occurs to the states labelled by four-row Young tableaux
$[m_1, m_2, m_3, m_4]$. The contributions to $M_z$ of both spin
and orbital are positive in this regime. The second regime
corresponds to the level crossing between the state labelled by
three-row Young tableaux, i.e., $m_4=0$. Because the contribution
of spin to $M_z$ is positive, but that of orbital to $M_z$ is
negative, we call it the regime of strong spin polarization and
orbital anti-polarization (SSP-OAP) regime. The third regime is
related to the level crossing occurring among the states labelled
by two-row Young tableaux ($m_3=m_4=0$) where spin degree of
freedom is frozen and orbital degree of freedom takes part in the
polarization process only. We call it the regime of spin frozen
and orbital polarization (SF-OP). Eventually, the magnetization
reaches saturation and then both the spin and orbital become
frozen (SF-OF) which is the phase IV indicated in Fig.
\ref{FIGURE_PHASE}. For $g_t=1.0$ the three critical values of the
magnetic field are $H^c_1=0.31$, $H^c_2=0.93$, and $H^c_3=4.0$.

Define $R=g_t/g_s$, we plot the phase diagram of $H$ and $R$ in
Fig. \ref{FIGURE_PHASE}. The critical lines that separate the
aforementioned regimes are plotted in Fig.(\ref{FIGURE_PHASE}).
Both the spin and orbital degree of freedom are frozen in the
phase IV. There is a simple relation for the boundary line between
SF-OP and SF-OF regimes, $R=2/H$, which arises from the
competition between the state related to the Young tableau $[N-1,
1]$ and $[N]$. The state in SSP-WP regime are quantum liquid of
spin-orbital mixture while SF-OP is orbital quantum liquid. If
$R=0.0$, the state in SSP-WOP has SU(2)$\otimes$U(1) symmetry,
while that in SF-OP has SU(2) symmetry \cite{YYamashita00}. When
$R$ approaches to unit, however, the third phase SF-OP disappears.
the phase SSP-OAP directly transits to phase SF-OF at point
$H^c=2.0$ then it undergoes only three distinct regimes which will
be explained again in the following.

In the above we interpreted the phases with the help of the
properties of Young tableaux. It can also be understood on the
basis of mean-field analysis.  In the presence of magnetic field,
at first, the Zeemann term in the Hamiltonian brings about three
processes: orbital flipping up, spin flipping up or both spin and
orbital flipping up simultaneously, which is the phase I indicated
in Fig. \ref{FIGURE_PHASE}. When the magnetic field exceeds
certain value for $0<g_t<g_s$, the three flipping processes,
$|\overline\downarrow\rangle\rightarrow|\underline\downarrow\rangle$,
$|\overline\downarrow\rangle\rightarrow|\overline\uparrow\rangle$
and
$|\overline\downarrow\rangle\rightarrow|\underline\uparrow\rangle$
are stopped. In phase II, the following flipping processes take
place,
$|\underline\downarrow\rangle\rightarrow|\underline\uparrow\rangle$,
$|\overline\uparrow\rangle\rightarrow|\underline\uparrow\rangle$
and
$|\underline\downarrow\rangle\rightarrow|\overline\uparrow\rangle$.
The last one is a process of spin flipping together with orbital
anti-flipping, which suppress the susceptibility (slop of the
magnetization curve) in phase II in comparison to that in phase I.
As is known \cite{YQLi99} that the ground state of the present
model in the absence of exterior field is an SU(4) singlet whose
physics picture can be considered as the simultaneous singlet of
the {\it spin} and {\it orbital}, as well as the {\it product of
spin and orbital}. Starting from the Ne\'el state of the {\it
product of spin and orbital}, the mean-field analysis tell us that
a pair of parallel spins will favor their orbital to be
anti-parallel, or vice versa. This makes it easy for us to
understand the spin flipping but orbital anti-flipping process in
phase II. Increasing the magnetic field will make it cross to the
edge of phase II where the spins are completely polarized. Thus in
phase III, the orbital start to flip back
$|\overline\uparrow\rangle\rightarrow|\underline\uparrow\rangle$.
Once all orbital become completely polarized it goes into phase
IV.

If $g_t=0$, the phase I transits to phase III directly. Because
the orbital SU(2) is not broken, orbital up and down are
equivalent and always half were in orbital up and half in orbital
down among the spin flipping process in phase I. Increasing the
magnetic field until all spins are completely polarized, the
orbital liquid state still stay in a liquid state and it will not
flip even if increasing the magnetic field further. So it will not
go into phase IV.  If $g_t=g_s$, the pure spin flipping and
orbital flipping have the same probability in phase I. In phase II
the spin flipping with orbital anti-flipping dose not occur and
pure spin flipping and orbital flipping have still the same
probability. Thus spin polarization itself will not happen unless
spin and orbital are completely polarized simultaneously.
Therefore it transits from phase II into phase IV directly.

In summary, we have investigated the magnetic properties of SU(4)
spin-orbital chains by numerical solutions of their Bethe-ansatz
equations. If the exterior field is decoupled to the orbital
degree of freedom, i.e. $g_t=0$, we found that there exists a
critical point of the quantum phase transition which separates it
into two regimes. If the magnetic contribution of the orbital is
not neglected, however, there will be three critical points and
four regimes that demonstrate different magnetic properties.
Though our calculation is made in one dimension, the analysis
based on the symmetry and symmetry breaking has no restrictions in
dimension. For a high dimensional systems (two or three
dimensional), it is speculated to have similar features.

This work is supported by trans-century projects and Cheung Kong
projects of the China Education Ministry. Helpful discussions with
F.C. Zhang and K. Yang are also acknowledged.


%

\end{multicols}
\end{document}